# Organic nanowires and chiral patterns of tetracyanoquinodimethane (TCNQ) grown by vacuum vapor deposition

J. C. Li,[a)] and Z. Q. Xue

*Department of Electronics, Peking University, Beijing 100871, People's Republic of China*

(Date 04/18/2005)

Organic nanowires and quasi-two-dimensional chiral patterns of tetracyanoquinodimethane have been successfully generated by vacuum thermal evaporation. The nanowires and patterns were characterized by using atomic force microscopy and transmission electron microscopy. The influence of electric charged clusters, deposition rate, and substrate temperature were experimentally investigated. Contrary to previous reports, charged clusters are found to be unnecessary to the chiral pattern formation. It was shown that the nanowire and pattern formation should be mainly dominated by its special crystallization properties, though the effect of the growth conditions cannot be neglected.

Nanoscale materials, such as carbon nanotubes,[1–3] nanowires[4–6] and nanocables,[7] have attracted increased attention due to their valuable electrical and optical properties, as well as their potential application to nanoscale devices.[8–9] However, most of these materials, either semiconductors or metals, are incompatible with the working medium, i.e., polymers or smaller organic molecules, required by potentially cheap and flexible organic electronic devices.[10–13] Though organic nanoscale tubes have been synthesized from larger organic molecule solutions by the self-assembling method,[14-15] its 'wet' procedure is unfavorable to many applications of nanoscale devices. Hence, it would be a challenge to find some 'dry' method, e.g., vacuum evaporation, to fabricate organic nanoscale materials, especially organic nanowires with desirable properties.

As a smaller organic molecule with 'conjugated' structure, tetracyanoquino-dimethane (TCNQ) has been extensively studied in past decades due to its extraordinary electrical and optical properties.[16–19] Recently, it was reported the observation of quasi-two-dimensional chiral patterns in growth of $C_{60}$-TCNQ by an ionized-cluster-beam (ICB) method.[20] Theoretical investigations have also been carried out on the broken symmetry of these patterns,[21] where, motivated by the presence of electric charged clusters in ICB deposition, the charged clusters are considered to be the crucial role that determines the chiral pattern formation. However, the actual growth mechanism and the role of charged clusters are still under debate. It would be necessary to establish a clear correlation between the pattern formation and the charged clusters. Using some growth methods other than ICB technique, without involving charged components, might be a solution.

In this letter, we show the observation of TCNQ nanowires and a series of striking patterns, from achiral to chiral, generated by vacuum thermal evaporation. By using this simple method, we are able to exclude the influence of charged clusters on the pattern formation. The effect of other growth factors is also studied.

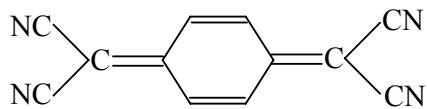

Fig. 1  Molecular structure of TCNQ.

Highly pure TCNQ material was purchased from the Sigma Chemical Co. The molecular structure of TCNQ is shown as Fig. 1. Firstly, put some TCNQ into a glass crucible (5-mm-inner-diameter and 35-mm-height). Then, set the crucible into a tungsten wire basket located in a vacuum chamber in which there is a shelter between the sample holder and the crucible. The distance between the mouth of the crucible and the substrate is 50 mm. The base vacuum of the deposition system was $1.0 \times 10^{-5}$ Torr. During evaporation, the pressure of the vacuum chamber was about $3.0 \times 10^{-5}$

---

[a)] Present address: Department of Chemistry, The University of Chicago.

Torr. Atomic force microscopy (AFM) samples, prepared on 150-nm-Au coated silicon, were observed using a NT-MDT P47 in tapping mode at room temperature in ambient conditions. Transmission electron microscopy (TEM) characterization was carried out using a JEM-200CX instrument, with samples fabricated on Cu grid pre-coated with amorphous carbon films. All the images presented here have not been processed in any way. Without special indication, the substrate is kept at room temperature.

Figure 2 shows a series of AFM images of the typical TCNQ nanowires. In general overview, AFM images showed that the obtained products are some randomly oriented rod-like structures with about 500-nm-diameter and more than 1000-nm-length (see Fig. 2(a)). However, as indicated by arrows in Fig. 2(b) and 2(c), AFM images of individuals revealed that these 'rods' were actually some three-dimensional TCNQ nanowire bundles. Each bundle consists of 15−30 closely parallel assembled nanowires. The wire diameter ranges from 30 to 80 nm, the wire head tends to be round-shaped.

Similar bundling phenomena have been reported in growth of single-wall carbon nanotubes[2] and Si nanowires.[4] As that of single-wall carbon nanotubes, bounding of TCNQ nanowires should occur during the growth and every bundle should be a packed crystal. Further experiments showed that the substrate morphologies have some effect on the wire diameter and length.[3] Whereas, no analogous nanowires were found in our study of polyaniline–TCNQ complex films[22] or other organic materials derived from TCNQ using the same method.[23–24]

Compared with the mentioned nanoscale materials, the growth of TCNQ nanowires doesn't need catalysts, complicated promoters or high temperature. In view of their small diameter and large aspect ratio, the fundamental properties of the TCNQ nanowires should be very different to those of their bulk forms. As we know, TCNQ is a smaller symmetry molecule with a 'conjugated' structure, which tends to form parallel molecular stacks. The bonding energy between neighboring molecules in the stack is much stronger than that between neighboring stacks. As a result, electrons can move easily along the stacks, but not between them.[13] As yet, we have no clear concept on the properties of individual TCNQ nanowire. Nevertheless, it would be a creative work to make hybrid or complex TCNQ nanowires, as it has been realized in some TCNQ bulk materials,[18–19] so as to endow them with novel electric and/or optical properties.

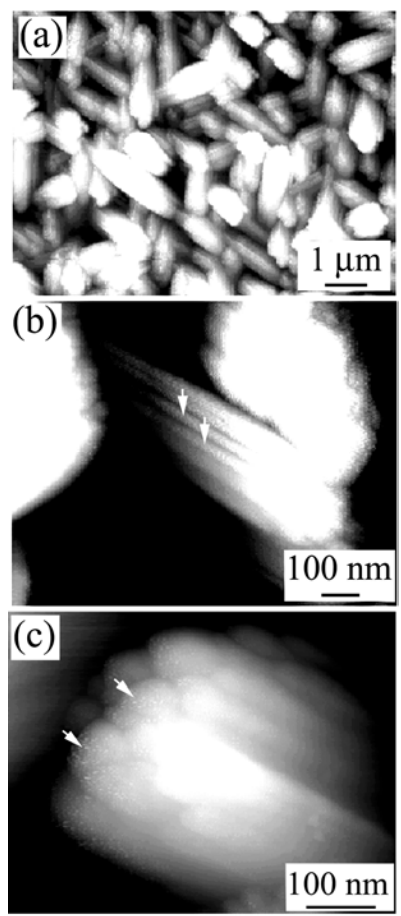

Fig. 2 Tapping mode AFM images of TCNQ nanowires: (a) Uniform TCNQ nanowire bundles with about 500-nm-diameter and more than 1000-nm-length. (b) Parallel closely packed nanowires of a bundle, with two of them indicated by arrows. (c) Another bundle, with arrows indicating the round-shaped head of the nanowires.

As shown in Fig. 3, typical TCNQ patterns, from achiral to chiral, were observed in TEM examination. Here, Fig. 3(a-c) were observed from different holes of Cu grid 1[#], while Fig. 3(d) was imaged from grid 2[#]. The two samples, prepared simultaneously, were placed at different position of the sample holder. The deposition rate of sample 2[#], placed at the center, was higher than that of sample 1[#] placed at the border. It was noted clearly that there exist obvious difference in the chirality of the patterns among different holes of the same grid, but almost identical in the same hole. Note that the pattern size of sample 2[#] is about three times larger than that of grid 1[#]. There exists a characteristic scale in the pattern growth, and when the pattern reaches the utmost size, the later coming clusters prefer to assemble to the center of the patterns and thus form a 'wing'-like structure (see Fig. 3(d)). These results should imply that the chiral pattern formation have something to do with the growth conditions, such as substrate microstructure, depositing rate and so on.

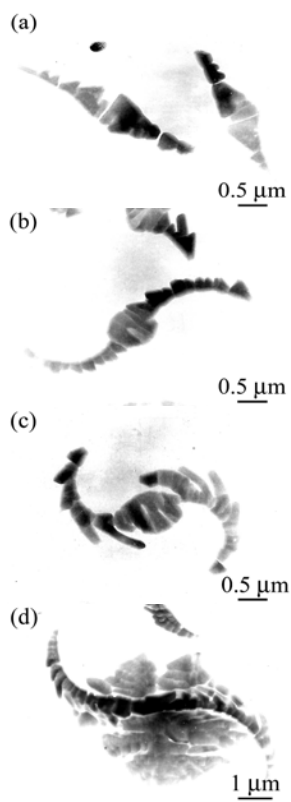

Fig. 3 TEM images of the typical quasi-two-dimensional TCNQ patterns which are all polycrystalline: (a) Achiral pattern with un-curved arms. (b) Chiral pattern with less curved arms. (c) Chiral pattern with long fins on the outer edge of the strong curved arms. (d) Characteristic scaled chiral pattern with 'wings'.

Interesting phenomena were observed when the substrate temperature is increased to 70 °C during evaporation. As shown in Fig. 4(a), TEM image indicates that the chiral patterns prefer to parallel packed together by forming a piece of crystal at center. Note that each such 'group' consists of 3 to 6 patterns with the same chirality, which is very similar to that of the correlated helix angles of SWNTs within a bundle.[2] Fig. 4(b) shows an unperfected example, where all the patterns failed to form their counter main branches. In fact, this result revealed the unique crystallization process, from quasi-two-dimensional chiral pattern growth to three-dimensional crystal formation, of TCNQ materials.

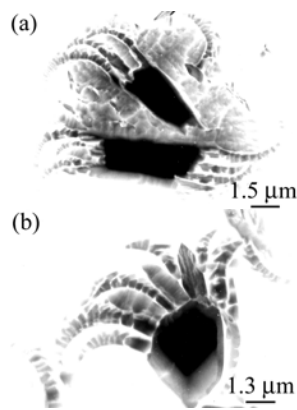

Fig. 4 Typical TEM images: (a) Assembled chiral patterns closely packed together from center by forming a piece of crystal. Each such 'group' consist of 3 to 6 patterns with same chirality. (b) An imperfect one with each pattern fail to grow out their counter main arms.

Unlike ICB technique, there is no electron beam, no ionizing voltage and no high accelerating voltage in our used method. In addition, the evaporating temperature of TCNQ materials is relatively too low to result in charged clusters. Moreover, TEM examinations of different samples, prepared simultaneously with or without applied electric field ($0–10^3$ V/cm), showed that there are no obvious difference among the patterns of them. Therefore, contrary to the reported electrostatic growth models,[21] we conclude that charged clusters are unnecessary to the chiral pattern formation.

A simple explanation is that these unusual growth phenomena should arise from some intrinsic properties of TCNQ itself, especially its special crystallization behavior. However, this does not imply that the growth factors, such as charged clusters, substrate microstructures and deposition rate etc., could be neglected. It is suggested that these factors should be considered to be the disturbances influencing on the pattern growth. Under ideal growth conditions, i.e., without disturbance, the as-obtained patterns should prefer to be uncurved (see Fig. 3(a)). More generally, the varying fabricating conditions are not ideal and thus result the familiar observed chiral patterns.

In summary, TCNQ nanowire bundles and striking patterns, from achiral to chiral, have been successfully generated by vacuum thermal evaporation. It was found that electric charged clusters are not necessary to the chiral pattern formation. The growth factors, such as deposition rate, substrate temperature and microstructures and so on, could affect the pattern growth to some extent. We suggest that the growth of TCNQ nanowire bundles and unique patterns should arise from its special crystallization behavior and even other unknown properties.

Financial support from the Doctoral Program Foundation of Education Ministry, People's Republic of China.

1. S. Iijima, Nature (London) **354**, 56 (1991).
2. A. Thess, R. Lee, P. Nikolaev, H. Dai, P. Petit, J. Robert, C. Xu, Y. H. Lee, S. G. Kim, A. G. Rincher, D. T. Colbert, G. E. Scuseria, D. Tomanek, J. E. Fischer, and R. E. Smalley, Science **273**, 483 (1996).
3. Y. C. Choi, Y. M. Shin, S. C. Lim, D. J. Bae, Y. H. Lee, B. S. Lee, and D. C. Chung, J. Appl. Phys. **88**, 4898 (2000).
4. B. Marsen and K. Sattler, Phys. Rev. B **60**, 11593 (2000).
5. P. A. Smith, C. D. Nordquist, T. N. Jackson, T. S. Mayer, B. R. Martin, J. Mbindyo, and T. E. Mallouk, Appl. Phys. Lett. **77**, 1399 (2000).
6. D. N. Mcilroy, D. Q. Zhang, R. M. Cohen, J. Wharton, Y. Geng, M. G. Norton, G. D. Stasio, B. Gilbert, L. Perfetti, J. H. Streiff, B. Broocks, and J. L. Mchale, Phys. Rev. B **60**, 4874 (1999).
7. X. C. Wu, W. H. Song, B. Zhao, W. D. Huang, M. H. Pu, Y. P. Sun, and J. J. Du, Solid State Commun. **115**, 683 (2000).
8. S. W. Chung, J. Y. Yu, and J. R. Heath, Appl. Phys. Lett. **76**, 2068 (2000).
9. J. Lefebvre, R. D. Antonov, M. Radosavljevic, J. F. Lynch, M. Llaguno, and A. T. Johnson, Carbon **38**, 1745 (2000).
10. J. C. Scott, Science **278**, 2071 (1997).
11. C. J. Drury, C. M. J. Mutsaers, C. M. Hart, M. Matters, and D. M. de Leeuw, Appl. Phys. Lett. **73**, 108 (1998).
12. K. Ziemelis, Nature (London) **393**, 619 (1998).
13. D. Voss, Nature (London) **407**, 442 (2000).
14. M. R. Ghadiri, J. R. Granja, R. A. Milligan, D. E. McRee, and N. Khazanovich, Nature (London) **366**, 324 (1993).
15. A. Harada, J. Li, and M. Kamachi, Nature (London) **364**, 516 (1993).
16. D. S. Acker, R. J. Harder, W. R. Hertler, M. Mahler, L. R. Benson, and W. E. Mochel, J. Am. Chem. Soc. **82**, 6408 (1960).
17. W. D. Grobman, R. A. Pollak, D. E. Eastman, E. T. Maas, B. A. Scott, Phys. Rev. Lett. **32**, 534 (1974).
18. W. E. Broderick, J. A. Thompson, E. P. Day, and B. M. Hoffman, Science **249**, 401 (1990).
19. R. Kumai, Y. Okimoto, and Y. Tokura, Science **284**, 1645 (1999).
20. H. J. Gao, Z. Q. Xue, K. Z. Wang, Q. D. Wu, and S. J. Pang, Appl. Phys. Lett. **68**, 2192 (1996).
21. I. M. Sandler, G. S. Canright, H. J. Gao, S. J. Pang, Z. Q. Xue, and Z. Y. Zhang, Phys. Rev. E **58**, 6015 (1998).
22. J. C. Li, Z. Q. Xue, Y. Zeng, W. M. Liu, Q. D. Wu, Y. L. Song, and L. Jiang, Thin Solid Films **374**, 59 (2000).
23. J. C. Li, Z. Q. Xue, X. L. Li, W. M. Liu, S. M. Hou, Y. L. Song, L. Jiang, D. B. Zhu, X. X. Bao, and Z. F. Liu, Appl. Phys. Lett. **76**, 2352 (2000).
24. J. C. Li, Z. Q. Xue, W. M. Liu, S. M. Hou, X. L. Li, and X. Y. Zhao, Phys. Lett. A **266**, 441 (2000).